\documentclass{jaa}
\usepackage{natbib}
\usepackage{xcolor}
\usepackage{graphicx}

\begin{document}\sloppy

\title{An alternative scheme to estimate AstroSat/LAXPC background for  faint sources}


\author{Ranjeev Misra\textsuperscript{1}, Jayashree Roy\textsuperscript{1} and J. S. Yadav\textsuperscript{2}}
\affilOne{\textsuperscript{1}Inter-University Centre for Astronomy and Astrophysics, Pune, 411007, India.\\}
\affilTwo{\textsuperscript{2}Department of Physics, Indian Institute of Technology, Kanpur, 208016, India\\}


\twocolumn[{

\maketitle

\corres{jayashree@iucaa.in}

\msinfo{}{}

\begin{abstract}
  
An alternative scheme is described to estimate the layer 1 LAXPC 20 background for faint sources where the source contribution to the 50-80 keV count rate is less than 0.25 counts/sec (15 milli-crabs or  $6 \times 10^{-11}$ ergs/s/cm$^2$).  We consider 12 blank sky observations and based on their 50-80 keV count rate in 100 second time-bins, generate four template spectra which are then used to estimate
the background spectrum and lightcurve for a given faint source observation.
The variance of the estimated background subtracted spectra for the 12 blank sky observations is taken as the energy dependent systematic uncertainty which 
 will dominate over the statistical one for exposures longer than 5 ksecs. The estimated 100 second time bin background lightcurve in the 4-20 keV band with a 3\% systematic error matches with the blank sky ones. The 4-20 keV spectrum can be constrained for  a source with flux $\gtrapprox 1$ milli-crab. Fractional r.m.s variability of 10\% can be determined for a $\sim 5$
 milli-crab source lightcurve binned at 100 seconds.
To illustrate the scheme, the lightcurves and spectra of three different blank sky observations, three AGN sources (Mrk 0926, Mrk 110, NGC 4593) and LMC X-1 are shown.

\end{abstract}


}]


\doinum{12.3456/s78910-011-012-3}
\artcitid{\#\#\#\#}
\volnum{000}
\year{0000}
\pgrange{1--}
\setcounter{page}{1}
\lp{1}

\section{Introduction}
The sensitivity of the LAXPC instrument \citep{antia} onboard AstroSat \citep{yadav16, pca} to extract spectral and long term temporal information of faint sources depends critically on how well the background of the instrument is characterized. 
  The background variation is primarily due to the changing response of the instrument to a varying local charged particle distribution. The cosmic X-ray background from the $\sim 0.25$ square degree field of view contributes less than 10\% of the observed background and hence its cosmic variance does not contribute significantly to the background variation. The standard method \citep{antia} to estimate the background involves blank sky spectra obtained as a function of latitude and longitude of the satellite. For a given science observation, a blank sky observation is chosen which is typically one that is closest in time to the science observation. The background is estimated as that
expected for the latitude and longitude covered during the science observation based on the blank sky observation after taking into account gain variation between the blank sky and source observations. This method provides background spectra which differ from the true background by  roughly 3\% and has been extensively used for science analysis. This systematic uncertainty exceeds the statistical one for exposures longer than 5 ksecs.

It is prudent to have different and independent methods to estimate the background to provide confidence on the scientific results obtained. Here we describe such a scheme which assumes that for faint sources the detected flux in the high energy band (50-80 keV) can be attributed
to the background alone and hence can act as a proxy to measure the background level as a function of time. As shown in this work, the technique is applicable to sources that contribute less than 15 milli-Crab of flux in the 50-80 keV band.
The scheme named as ``Faint source background estimation'' has been incorporated in the LAXPC software laxpcsoft available at the AstroSat Science Support Cell\footnote{http://astrosat-ssc.iucaa.in/?q=laxpcData}. It has been used for scientific analysis of several faint sources ( e.g. LMC X-1 \citep{mudambi}, RGB J0710+591 \citep{pranju,yadav}).

\section{Estimating LAXPC background}
\begin{figure}
\begin{center}
{\includegraphics[width=.7\linewidth,angle=-90]{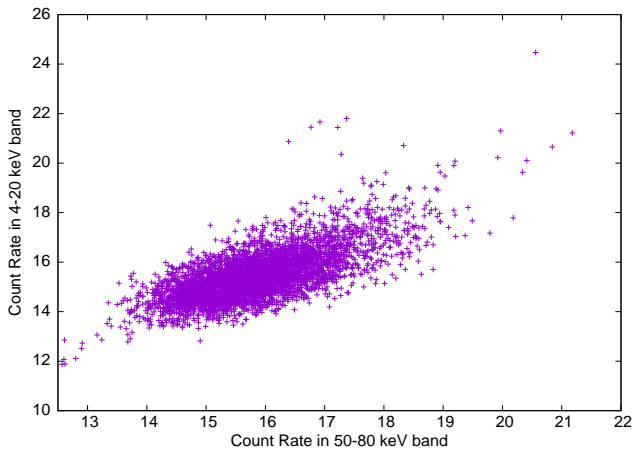}}
\end{center}
\caption{Count rates for the 4-20 keV and 50-80 keV plotted against each other
for LAXPC 20 Layer 1. The time-scale for integration is 100s.}
\label{countrate}
\end{figure}

\begin{figure}
  
\begin{center}
{\includegraphics[width=.7\linewidth,angle=-90]{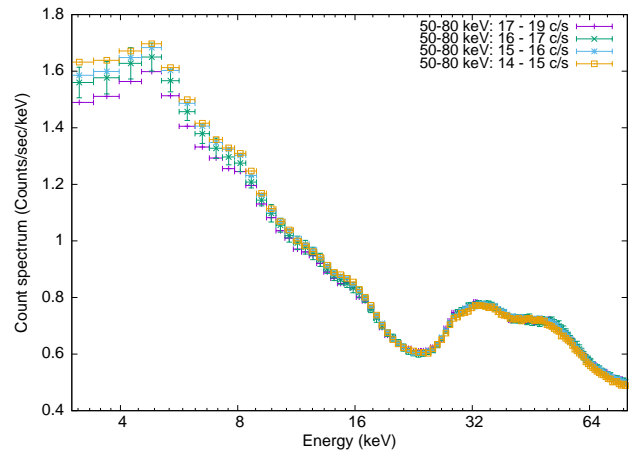}}
\end{center}
\caption{Average Blank sky spectrum corresponding to when the 50-80 keV count
  rates are in the range 14-15,15-16,16-17 and 17-20 counts/sec. The spectra
  have been normalized such that 50-80 keV flux levels are nearly equal, in
  order to highlight the different spectral shapes at low energies. These four
spectra are used as templates to estimate the background spectrum and lightcurve of a source observation.}
\label{segspec}
\end{figure}

We consider twelve blank sky observations that are listed in Table \ref{Table1}. For these observations, the lightcurve in 100 seconds for LAXPC 20 layer 1, were computed in energy bands 4-20 keV and 50-80 keV bands. The count rates of these two energy bands are plotted against each other in Figure \ref{countrate}. Most of the data lie within 50-80 keV count rate of 14 to 19 counts/sec.
We find that it is prudent to consider data only in this range, and divide the range into four parts corresponding to 14-15, 15-16, 16-17 and 17-19 counts/sec. The average spectra corresponding to these selections are shown in Figure \ref{segspec}. The spectra have been normalized such that 50-80 keV flux levels are nearly equal, in order to highlight the different spectral shapes at low energies\footnote{Due to gain variation, the channel to energy conversion for the blank sky observations are different. Here, we consider one observation spectrum as the reference and we interpolate the other spectra such that all of them have the same energy bin channels as that of the reference.}. We note that the spectral shapes are different at low energies and treat these four spectra as templates for estimating the background spectrum and lightcurve for a source.

  The procedure for estimating the background for a source is as follows:

  \noindent 1) Collect the 50-80 keV lightcurve in 100 second time bins.
  
  \noindent 2) Select  GTIs based on the count rate of the 50-80 keV energy range being between 14 and 19 counts/s.

  \noindent 3) For each time bin of the 50-80 keV, use the observed count rate to estimate the complete background energy spectrum using the four templates. The four templates are assigned to the mid point of their count rate in 50-80 keV, i.e. 14.5, 15.5, 15.5 and 18 c/s. The templates are then interpolated to obtain the corresponding spectrum appropriate for the observed count rate.

  \noindent 4) Integrate the estimated background energy spectrum for each time bin over the desired energy band.

  \noindent 5) Combine the estimated background energy spectra to estimate the time averaged background spectrum.

\begin{figure}
  
\begin{center}
{\includegraphics[width=.7\linewidth,angle=-90]{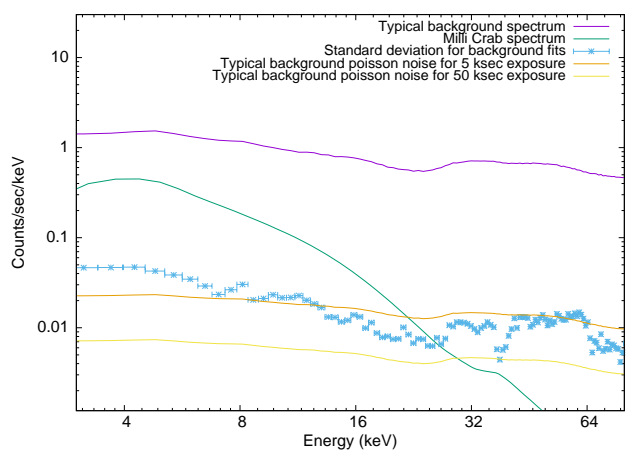}}
\end{center}
\caption{The standard deviation of the estimated background as compared to
  blank sky observations. Also shown for comparison are a typical blank sky
  spectrum, the spectrum for a 1 milli-crab source and the typical Poisson level
for the blank sky spectrum. }
\label{stdspec}
\end{figure}

  To test the efficacy of the method, the background spectra was estimated for each of the blank sky observations using the method described above. The estimated background was then compared with the observed spectrum and at each energy bin. The twelve blank sky observations were used to get the standard deviation of the estimated background and the observed spectra as a function of energy. The standard deviation in units of counts/sec/keV are shown as a function of energy in Figure \ref{stdspec}. This standard deviation can be used as the energy dependent systematic error on an estimated background spectrum obtained from this method.  
  The standard deviation is compared with a typical background spectrum shown in Figure \ref{stdspec} and hence the systematic error on the background is of the order of a few percent. Thus, the systematics attained from this method are of the same order as that from the standard technique. The standard deviation of the background subtracted count rates in 4-20 keV band for the 12 blank sky observations is around 0.3 c/s, indicating a 3-sigma detection of a source to be $\sim 1$ c/s in this energy band. 
  Also shown is the 1 milli-crab source spectrum, which reveals that although the source count rate is a  factor of few below the background, it should be detected using this method. The systematics at 25 keV is comparable to the source flux from a 1 mCrab source. For comparative reference, the typical Poisson noise levels for an exposure of 5 and 50 ksecs are shown. Note that the systematic error dominates over the Poisson one for exposures longer than 5 ksecs. The software includes this error in the background
  spectrum file. 

  The estimated background lightcurve in any energy band is based on the count rate in the 50 -80 keV band. The deviation of the estimated
    lightcurve from the true background is due to the systematic limitations of the technique
    and the typical Poisson error of the 50-80 keV band count rate in 100 second time bin which is  $\sim 2.5$\%. As shown in the next section, an addition of a uniform 3\% error on the estimated lightcurve at each
  100 second time bin, leads to consistent estimates of the expected variance for blank sky observations.

  The spectral templates used in this technique correspond to 50-80 keV count rates separated by 1 count/sec and which is assumed to be only from the background. Hence this scheme is limited to cases when the source count rate
  is significantly less than the template separation rate. Thus it is applicable for sources with count rate $< 0.25$ counts/s  which  translates to 15 milli-crabs or  $6 \times 10^{-11}$ ergs/s/cm$^2$ in the 50 - 80 keV band.

The estimated background lightcurves and spectra for LAXPC 10 can also be estimated using the same technique. However, since
LAXPC 10 has an higher background with larger uncertainty than LAXPC 20, it is recommended that LAXPC 20 should be primarily used for such analysis and LAXPC 10 results to be taken as a corroboration.  

\section{Verification and Examples}

To illustrate the method, lightcurves and spectra were generated for three blank sky observations from 2017, 2018 and 2019 which were not part
of the observations used to obtain the templates, three Active Galactic Nuclei (Mrk 0926, Mrk 110 and NGC 4593), and for the extra-galactic X-ray binary LMC X-1. The lightcurves were generated for a time bin of 100s and for 4-20 keV energy range.

The left panels of Figures 4 and 5, shows the total lightcurve (i.e. source with background marked as Src+Bkg), the estimated background
lightcurve and the subtraction of the two for the blank sky observations from 2017 and 2019. The 2019 blank sky observation shows increased count rate for two times just before the satellite entered the SAA. If the two increases are removed from the data then the resultant lightcurves are similar to ones obtained for the 2017 data (middle panel of Figure \ref{figbkg19}).
While the reason for these higher counts rates is not clear,
such variations just before entry into SAA should be treated with caution
in science analysis. These are most likely due to local variation/fluctuations in geomagnetic field. This is to emphasize that proper time-selection by inspection is required to obtain
reliable results. The right panels of Figures 4 and 5, show
the residuals of the observed spectra over the background for the
2017 and 2019 blank sky observations. Although the residuals are shown for a wide energy band,
note that the technique is only valid in 4-30 keV range. This shows that the systematics included in the background spectra are adequate and no significant residuals are seen in
the 4-30 keV band. Residuals of the order of 0.02 counts/sec/keV should be attributed
to systematics. Since a 3\% uncertainty has been included
  in the estimated background lightcurve, the error on the background subtracted lightcurve
  is a combination of the Poisson noise in the observed lightcurve and this background uncertainty.
Table \ref{Table2} lists the average count rate, the variance ($\sigma^2$), the
standard deviation ($\sqrt(\sigma^2)$), the expected variance ($\sigma_{EV}^2$) and the expected standard deviation ($\sqrt(\sigma_{EV}^2)$) for the background
subtracted lightcurves binned in 100 and 1000 seconds. For these three blank sky observations the average count rate is within the 0.3 c/s deviation found for the twelve blank sky used as the template for the technique. The standard deviation
($\sqrt(\sigma^2)$) and the expected one ($\sqrt(\sigma_{EV}^2)$) are similar.
This implies that the
background estimation technique can be applied to study lightcurves in 100 to 1000 second
binning over a time-span of $\sim 40$ ksecs. Note that the errors quoted in the Table
 \ref{Table2} for the measured and expected variance are statistical ones and hence maybe underestimated.
The recommendation is that an excess variance of more than 1 c/s, should be considered as evidence of variability.

 The background subtracted
spectra for the AGN sources were fitted using a power-law and for LMC X-1
a disk emission model and power-law was used. Figure 6  shows the
background subtracted spectrum along with the expected background spectrum (top
panel) and residuals (bottom panel) for Mrk 0926. We emphasize that the spectra here are
shown for the full energy range of 4-80 keV for illustration, and
science analysis should be limited to 30 keV, since the higher energy 50-80 keV
spectrum has been used to calibrate the background. Similar results were
obtained for the other sources and the residuals
are of the order of 0.02 counts/sec/keV as expected from the blank sky observations. 

Table \ref{Table2} lists the properties of the background subtracted lightcurves
for the sources and the flux in the 4-20 keV band. Mrk 0926 and Mrk 110 are clearly detected with a background
subtracted average count rate of 8 and 6 c/s (Table \ref{Table2}), but show no
variability with the observed standard deviation of the same order as
the expected one. On the other hand NGC 4593 has a lower
count rate of 4 c/s but shows slight evidence of variability with a r.m.s
of $\sqrt(\sigma^2-\sigma_{EV}^2) \sim 0.7$ c/s.
The AstroSat observation of extra galactic source LMC X-1 has
been reported by \citet{mudambi}. It shows a count rate of 19 c/s with clear evidence of variability with r.m.s of $\sim 1.2$ c/s. These results show that a 
LAXPC observation of a source with a 4-20 keV band flux $\gtrapprox 2 \times 10^{-11}$ ergs/s/cm$^2$ (i.e. $\gtrapprox 2$ c/s or $\gtrapprox 1$ milli-crab), will
be able to constrain the source spectrum.  For 100 second binned lightcurves
a variability  of $\sim 1.0$ c/s   can be detected  which translates to
a fractional r.m.s of 10\% for a $5$ milli-crab source.

\section{Summary and Discussion}

We have presented an alternate scheme to estimate the background
spectrum and lightcurve for AstroSat LAXPC 20 based on using the
detected count rate in the 50-80 keV as a measure of the background.
A software that incorporates the scheme is available at AstroSat Science Support Cell\footnote{http://astrosat-ssc.iucaa.in/?q=laxpcData}. The software also can compute the estimated background
lightcurves and spectra for LAXPC 10 using the same technique. However, since
LAXPC 10 has an higher background with larger uncertainty than LAXPC 20, it
is recommended that LAXPC 20 should be primarily used for such analysis and
LAXPC 10 results to be taken as a corroboration.

The scheme can be perhaps be improved by exploring possibilities to
better estimate the background rate. This includes considering only
single events to see if the correlation between the low and high energy
counts in the blank sky spectra is tighter. The correlation between
the low and high energy count rate does vary for different blank
sky observations and one can examine if any   
satellite system parameter can be used to predict the variation.
A related idea would be to study the correlation as a function of
latitude and longitude to see if there is any predictable trend.
These improvements may lead to a background estimation closer to the
Poisson limit for a 30 ksecs exposure.

The scheme is applicable to faint sources where the source contribution
to the 50-80 keV count rate is
less than 0.25 counts/sec (15 milli-crabs or  $6\times10^{-11}$ ergs/s/cm$^2$) and is limited
to the energy range 4-30 keV.
The systematic uncertainty in the background spectra will dominate over the statistical error for exposures larger than 5 ksecs and is of the same order as that from the standard technique. Thus, it will be prudent to confirm spectral results using both techniques.
The technique allows for background lightcurve estimation on time-scales larger than 100 seconds. The spectrum of a $\sim 1$ milli-crab source (i.e. 4-20 keV band flux of $\gtrapprox 2 \times 10^{-11}$ ergs/s/cm$^2$) can be constrained by LAXPC observations. Since the systematics dominate the Poisson statistics for exposures greater than 5 ksecs, the sensitivity of the instrument to measure the spectrum of a source does not improve for exposures longer than $\sim 30$ ksecs. Variability of a lightcurve binned at 100 seconds can be detected for a level greater than $1$ c/s in the 4-20 keV band, which translates to 10\% fractional r.m.s of a $5$ milli-crab source. Since the background lightcurve is estimated using the observed variability seen in the high energy band, it is expected to more reliable than the standard technique. Thus, using this technique, LAXPC can be used to study both the spectral and temporal properties of sources with flux greater than $5$ milli-crab.

\begin{table*}
\scriptsize
\caption{Details of the twelve blank sky observations used to generate the template spectra}
\centering
\begin{tabular}{llllll}
\hline
Target&Observation&R.A. &Decl.&Date&Exposure\\
      &ID	  &(deg)&(deg)&    &(ksecs)     \\
\hline\\
Sky-9\_75\_50	&G05\_156T09\_9000000604		&237.37	&47.10	&2016 Aug 16&32.0\\
Sky-5		&T01\_132T01\_9000000636		&57.37	&-47.10	&2016 Aug 30&27.4\\
Sky-6		&C01\_015T01\_9000000668		&7.65	&12.55	&2016 Sep 15&43.0\\
Sky-10		&G06\_115T01\_9000000734		&321.22	&-48.68	&2016 Oct 16&35.9\\
Sky-6		&C02\_011T01\_9000000850		&7.65	&12.55	&2016 Dec 03&39.4\\
Sky-3		&C02\_003T01\_9000000924		&129.48	&-27.89	&2016 Dec 24&32.1\\
Abell3535	&A02\_108T01\_9000001024		&194.45	&-28.49	&2017 Feb 11&49.1\\
Sky-9\_75\_50	&G07\_044T09\_9000001334		&237.37	&47.10	&2017 Jun 24&32.9\\
Sky\_4u1626	&G07\_049T02\_9000001354		&250.00	&-70.00	&2017 Jul 04&22.4\\
Sky-8		&C02\_021T01\_9000001482		&237.39	&70.35	&2017 Aug 21&39.5\\
Sky-9\_75\_50	&G08\_046T09\_9000001600		&237.37	&47.10	&2017 Oct 11 &28.9\\
Blank Sky 5 255-50&A04\_198T01\_9000001708		&57.37	&-47.10	&2017 Nov 21&35.6\\

\hline
\label{Table1}
\end{tabular}
\end{table*}

\begin{table*}
\tiny
\caption{The average counts, variance and expected variance of the background subtracted lightcurves and the flux in the 4-20 keV band for sources used to verify the scheme.}
\centering
\begin{tabular}{cccccccccccc}
\hline
Source&R.A. &Decl.&Date&Exposure&Time Bins& Average  & Variance ($\sigma^{2}$)&$\sqrt{\sigma}^{2}$ & Expected &$\sqrt{\sigma_{EV}}^{2}$ &Flux \\
&&&&&&&&&Variance ($\sigma_{EV}^{2}$)&&4-20 keV\\
&(deg)&(deg)&    &(ksec)&(sec)&c/s&$(c/s)^{2}$&$(c/s)$&$(c/s)^{2}$&$(c/s)$&$\times 10^{-11}$ergs $cm^{-2} s^{-1}$\\
\hline \hline
Blank Sky-8 &183.48&	22.80&	2017 Jan 10	&3.98 &100       &  -0.05$\pm$0.03   &0.40$\pm$0.03      &0.63$\pm$0.02      &0.36$\pm$0.03      &0.60$\pm$0.02& -     \\
(2017)      &	 &&&&1000      &  0.03 $\pm$0.04   &0.17$\pm$0.03      &0.42$\pm$0.04      &0.10$\pm$0.02      &0.31$\pm$0.03  &  -  \\
Blank Sky-8 &237.39&	70.35&	2018 Apr 11	&40.5 &100       &  -0.05$\pm$0.03   &0.24$\pm$0.02      &0.49$\pm$0.02      &0.40$\pm$0.03      &0.64$\pm$0.03 &   -  \\
(2018)      &	 &&&&1000      &  -0.02$\pm$0.05   &0.06$\pm$0.01      &0.25$\pm$0.02      &0.10$\pm$0.02     &0.32$\pm$0.03    & - \\
Blank Sky-8 &237.39&	70.35&	2019 Feb 01	&16.7&100       &  -0.16$\pm$0.05   &0.26$\pm$0.03      &0.51$\pm$0.03      &0.40$\pm$0.04      &0.62$\pm$0.04    &-  \\
(2019)      &	&&&&1000      &  -0.12$\pm$0.06   &0.10$\pm$0.03      &0.32$\pm$0.05      &0.09$\pm$0.03      &0.29$\pm$0.04    & - \\
Mrk 0926    &346.18&	-8.69&	2016 Nov 28	&9.9 &100       &  7.73 $\pm$0.07   &0.35$\pm$0.05      &0.59$\pm$0.04      &0.47$\pm$0.07      &0.69$\pm$0.05   & 6.6  \\
            &	 &&&&1000      &  7.64 $\pm$0.10   &0.06$\pm$0.02      &0.24$\pm$0.04      &0.16$\pm$0.06      &0.40$\pm$0.07     & \\
Mrk 110     &141.30&	52.29&	2017 Apr 15	&26.2&100       &  5.90 $\pm$0.04   &0.38$\pm$0.03      &0.62$\pm$0.03      &0.42$\pm$0.04      &0.65$\pm$0.03   & 4.9  \\
            &    &&&&1000      &  5.94 $\pm$0.06   &0.16$\pm$0.04      &0.40$\pm$0.05      &0.12$\pm$0.03      &0.35$\pm$0.04     & \\
LMC X-1     &84.91&	-69.74&	2016 Nov 25	&54.2 &100       &  18.31$\pm$0.03   &5.18$\pm$0.32      &2.28$\pm$0.07      &0.55$\pm$0.04      &0.74$\pm$0.02   & 10.1  \\
            &	 &&&&1000      &  18.42$\pm$0.04   &4.05$\pm$0.67      &2.01$\pm$0.17      &0.14$\pm$0.02     &0.37$\pm$0.03    &  \\
NGC 4593    &189.91&	-5.34&	2016 Jul 16	&39.9 &100       &  4.08$\pm$0.03    &0.96$\pm$0.07      &0.98$\pm$0.04      &0.40$\pm$0.03      &0.63$\pm$0.02    & 3.4 \\
            &	 &&&&1000      &  4.28$\pm$0.04    &1.04$\pm$0.20      &1.02$\pm$0.10      &0.10$\pm$0.02      &0.31$\pm$0.03   &  \\

\hline

\label{Table2}
\end{tabular}
\end{table*}

\begin{figure*}
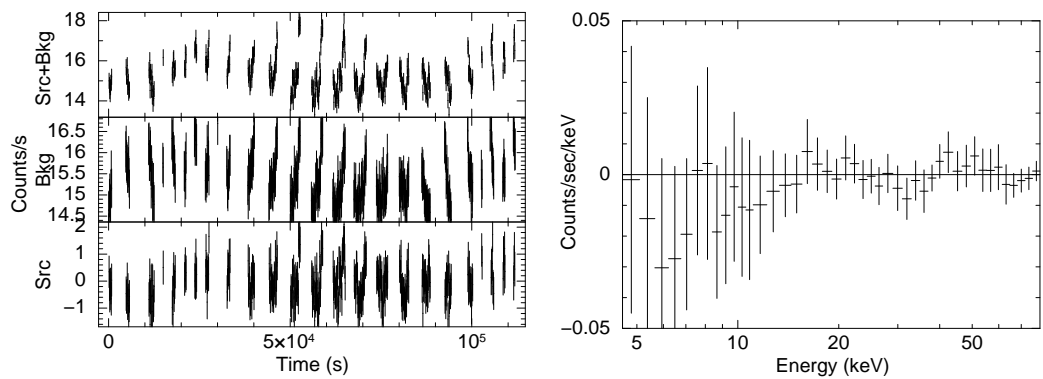

  \begin{center}
  {\includegraphics[width=.28\linewidth,angle=-90]{LC_BKG_2017_R2.eps}}
  {\includegraphics[width=.28\linewidth,angle=-90]{2017_bkg_R2.eps}}
  \end{center}
\caption{Left panels shows the lightcurves of the blank sky observations from 2017. Top panel shows the total source with background count rate, middle panel shows the background and the bottom panel shows the background subtracted count rate from the source. Blank spaces between the lightcurves pertains to the SAA passages. Right figure shows the residuals of the energy spectrum.  }
\label{figbkg17}
\end{figure*}

\begin{figure*}
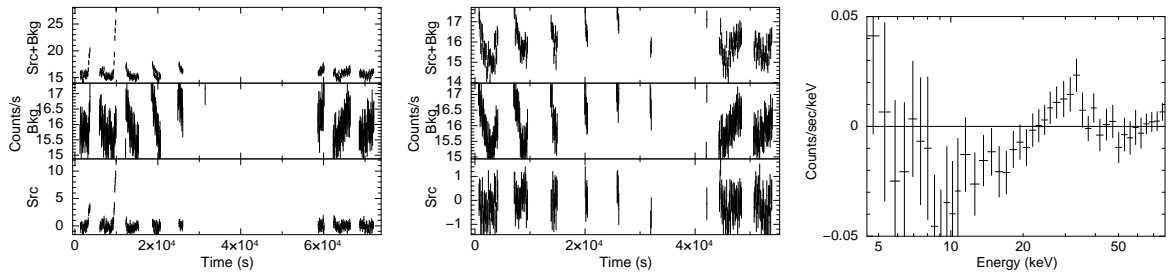

  \begin{center}
  {\includegraphics[width=.2\linewidth,angle=-90]{LC_BKG_2019_R2_all.eps}}
  {\includegraphics[width=.2\linewidth,angle=-90]{LC_BKG_2019_R2.eps}}
  {\includegraphics[width=.2\linewidth,angle=-90]{2019_bkg_R2.eps}}
  \end{center}
\caption{Left figure shows the lightcurves of the blank sky observations from 2019. Top panel of left figure shows the total source with background count rate, middle panel shows the background and the bottom panel shows the background subtracted count rate from the source. Middle figure shows the same lightcurves when the two increases due to entry of the satellite in SAA passages are removed from the data. Right figure shows the residuals of the energy spectrum.}
\label{figbkg19}
\end{figure*}

\begin{figure*}
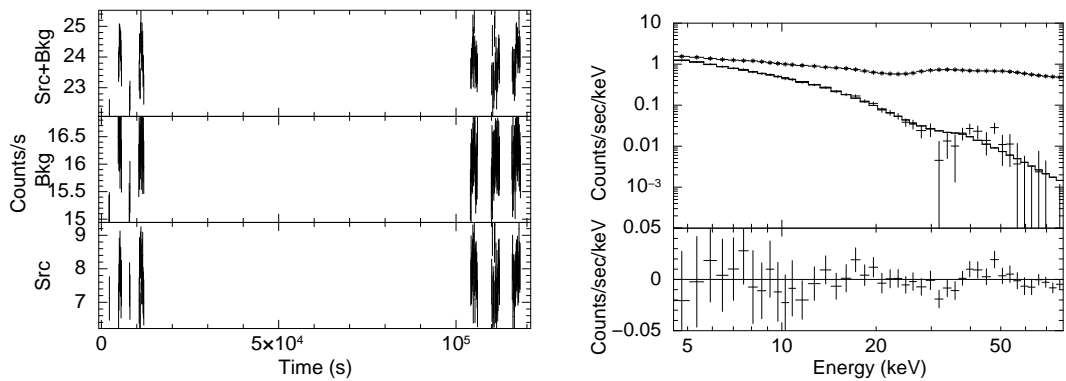

  \begin{center}
  {\includegraphics[width=.28\linewidth,angle=-90]{LC_MRK0926_R2.eps}}
  {\includegraphics[width=.28\linewidth,angle=-90]{MRK0926_spect_R2.eps}}
\end{center}
\caption{Left panel shows the lightcurves of the source Mrk 0926 (Top panel: total source with background count rate, Middle panel: the background and Bottom panel: the background subtracted count rate). Right panel shows the energy spectrum along with the expected background spectrum (top) and residuals (bottom) after  fitting with a power-law model.}
\label{fig926}
\end{figure*}

\section*{Acknowledgements}
This publication uses the data from the AstroSat mission of the Indian Space Research Organization (ISRO),
archived at the Indian Space Science Data Centre (ISSDC). We thank members of LAXPC instrument team for their contribution
to the development of the LAXPC instrument. This research has made use of software
provided by the High Energy Astrophysics Science Archive Research Center (HEASARC), which is a service of the Astrophysics
Science Division at NASA/GSFC. The authors thank the referee, Keith M Jahoda, for suggestions and comments which substantially improved the manuscript.
\vspace{-1em}



\end{document}